\documentclass[a4paper,11pt]{article} 
\pdfoutput=1 % JINST
\usepackage{jinstpub} % JINST

\usepackage{caption}
\usepackage{subcaption}
\usepackage{mathrsfs}
\graphicspath{{figures/}{/plots}} 

\usepackage{xspace}

\newcommand {\ie}{\mbox{i.e.}\xspace}     %i.e.
\newcommand {\eg}{\mbox{e.g.}\xspace}     %e.g.

\title{3D silicon pixel detectors for the High-Luminosity LHC}

\author[a,1]{J.~Lange\note{Corresponding author.},} 
\author[b]{M.~Carulla Areste,}
\author[a]{E.~Cavallaro,}
\author[a]{F.~F\"{o}rster,}
\author[a,c]{S.~Grinstein,}
\author[a]{I.~L\'{o}pez Paz,}
\author[a]{M.~Manna,}
\author[b]{G.~Pellegrini,}
\author[b]{D.~Quirion,}
\author[a]{S.~Terzo,}
\author[a]{D.~V\'{a}zquez Furelos}

\affiliation[a]{Institut de F\'{i}sica d'Altes Energies (IFAE), The Barcelona Institute of Science and Technology (BIST), 08193 Bellaterra (Barcelona), Spain}

\affiliation[b]{Centro Nacional de Microelectronica (CNM-IMB-CSIC), Campus UAB, 08193 Bellaterra (Barcelona), Spain}

\affiliation[c]{Instituci\'{o} Catalana de Recerca i Estudis Avan\c{c}ats (ICREA), Pg. Llu\'{i}s Companys 23, 08010 Barcelona, Spain}

\emailAdd{joern.lange@cern.ch} %JINST

\abstract{ %JINST

3D silicon pixel detectors have been investigated as radiation-hard candidates for the innermost layers of the HL-LHC upgrade of the ATLAS pixel detector. 3D detectors are already in use today in the ATLAS IBL and AFP experiments. These are based on 50$\times$250\,$\mu$m$^{2}$ large pixels connected to the FE-I4 readout chip. Detectors of this generation were irradiated to HL-LHC fluences and demonstrated excellent radiation hardness with operational voltages as low as 180\,V and power dissipation of 12--15\,mW/cm$^2$ at a fluence of about $10^{16}\,n_{eq}$/cm$^2$, measured at -25$^{\circ}$C. Moreover, to cope with the higher occupancies expected at the HL-LHC, a first run of a new generation of 3D detectors designed for the HL-LHC was produced at CNM with small pixel sizes of 50$\times$50 and 25$\times$100\,$\mu$m$^{2}$, matched to the FE-I4 chip. They demonstrated a good performance in the laboratory and in beam tests with hit efficiencies of about 97\% at already 1--2\,V before irradiation.

}

\keywords{Particle tracking detectors, HL-LHC upgrade, 3D silicon pixel detectors} %

\begin{document}

\maketitle %JINST
\flushbottom %JINST 

\section{Introduction}
\label{sec:intro}

3D silicon detectors present a radiation-hard sensor technology, since the distance of the columnar electrodes, which penetrate the sensor bulk perpendicular to the surface, is decoupled from the device thickness and thus can be chosen to be significantly smaller than for the standard planar sensors that have the electrodes along the surface~\cite{bib:3D}. This represents advantages in terms of less trapping at radiation-induced defects and lower operational voltages, which translates into lower power dissipation after heavy irradiation. During the last years, the 3D technology has matured from basic R\&D to a production phase, and by now 3D pixel detectors are already used in high-energy physics particle detectors where superior radiation hardness is key: in the ATLAS Insertable B-Layer (IBL)~\cite{bib:IBLprototypes} and the ATLAS Forward Proton (AFP) detector~\cite{bib:AFP3D2, bib:AFPproduction}. This first generation of 3D pixel sensors with inter-electrode distances of 67\,$\mu$m at a sensor thickness of 230\,$\mu$m demonstrated a radiation hardness of at least up to $5\times10^{15}\,n_{eq}$/cm$^2$~\cite{bib:IBLprototypes}.

For the High-Luminosity upgrade of the Large Hadron Collider (HL-LHC)~\cite{bib:HL-LHC}, envisaged to start operation in 2026 and targeting a 10-fold increase of integrated luminosity, the radiation-hardness requirements are even more demanding~\cite{bib:HL-LHCfluences}. A fluence of 1--2$\times10^{16}\,n_{eq}$/cm$^2$ is expected for the innermost pixel layer of the ATLAS and CMS experiments at the end of life time after an integrated luminosity of 3,000\,fb$^{-1}$. Moreover, for occupancy reasons, smaller pixel sizes of 50$\times$50 or 25$\times$100\,$\mu$m$^{2}$ are planned.
%, in combination with a new radiation-hard readout chip developed by the CERN RD53 collaboration.

Due to their intrinsic radiation hardness, 3D pixel sensors are promising candidates for the innermost pixel layers for the HL-LHC upgrade of the ATLAS detector. In this work, their suitability for this is studied in two different steps. Firstly, the radiation hardness of the already existing IBL/AFP generation is investigated up to HL-LHC fluences. Secondly, a new dedicated HL-LHC generation of 3D sensors is developed and tested, which is designed for the small pixel sizes of the HL-LHC and to even further improve the radiation hardness with smaller inter-electrode distances.

\section{Radiation hardness of the IBL/AFP 3D generation at HL-LHC fluences}
\label{sec:IBL-AFPgeneration}

\subsection{Devices}
\begin{figure}[hbtp]
	\centering
	\includegraphics[width=15cm]{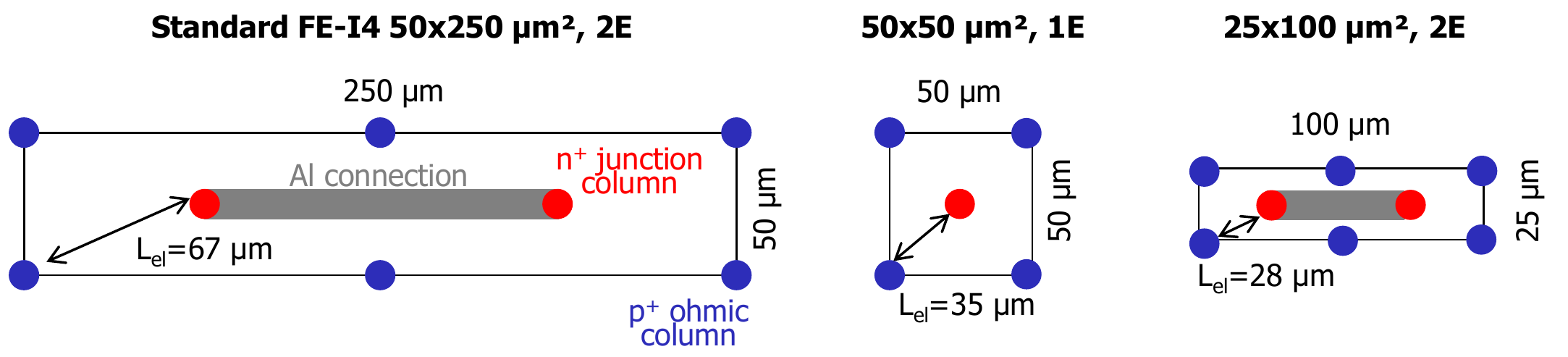}
	\caption{Geometry of a 3D pixel cell for a standard IBL/AFP FE-I4 pixel with 50$\times$250\,$\mu$m$^{2}$, 2E configuration (left), and for a 50$\times$50\,$\mu$m$^{2}$, 1E, (centre) and a 25$\times$100\,$\mu$m$^{2}$, 2E, (right) pixel.}
	\label{fig:3Dgeometry}
\end{figure}

The IBL/AFP-generation 3D pixel sensors have a pixel size of 50x250\,$\mu$m$^{2}$ with two 3D n$^+$~junction columns shorted together per pixel (2E configuration), giving an inter-electrode distance of $L_{el}$=67\,$\mu$m (see figure~\ref{fig:3Dgeometry}, left). The p-type substrate has a thickness of 230\,$\mu$m and a resistivity of about 20\,k$\Omega$cm. The sensors are compatible to the FE-I4 readout chip~\cite{bib:FEI4} with a 336$\times$80 pixel matrix, representing a total sensitive area of 16.8$\times$20.0\,mm$^2$. For testing purposes, also pixel sensors compatible with the ATLAS FE-I3 readout chip were used, which have 160$\times$18 pixels of 50$\times$400\,$\mu$m$^{2}$ size, with three n$^+$ junction columns per pixel (3E, $L_{el}$=71\,$\mu$m). Left-over sensors from the IBL production of CNM (Centro Nacional de Microelectronica, Barcelona, Spain)~\cite{bib:CNMIBLProduction} were used, which were bump-bonded to the readout chips and assembled on a readout board at IFAE Barcelona.

%In order to study the radiation hardness of the IBL/AFP 3D detectors beyond the established IBL level of $5\times10^{15}\,n_{eq}$/cm$^2$, irradiation campaigns with protons and neutrons were performed, and the performance of the pixel devices was measured in the laboratory and beam tests.

%\subsection{Non-uniformly proton-irradiated FE-I4 3D detectors}
\subsection{Irradiation, hit efficiency and power dissipation}

For the study of the IBL/AFP 3D generation beyond the established IBL level of $5\times10^{15}\,n_{eq}$/cm$^2$, the strategy was to use non-uniformly proton-irradiated FE-I4 devices for the determination of hit efficiencies and operation voltages in a beam test and uniformly neutron-irradiated FE-I3 devices for the determination of leakage current and power dissipation. This disentanglement was chosen for technical reasons: On the one hand, FE-I4 devices are needed for low thresholds of about 1.5\,ke$^-$, which is key for a good efficiency. But at that time they could be only non-uniformly irradiated with 24\,GeV protons at the CERN-PS IRRAD facility (lower energy protons at other facilities would have allowed uniform irradiation, but damaged the chip more). Uniform irradiation (ideal for the study of leakage current and power dissipation) was possible at JSI Ljubljana with neutrons. However, tantalum in the FE-I4 chip leads to a high activation during neutron irradiation, so that pixel devices with the FE-I3 chip were chosen instead (which in turn are not ideal for efficiency measurements due to minimal thresholds of about 3\,ke$^-$). 

\begin{figure}[hbtp]
	\centering
	\includegraphics[width=7.4cm]{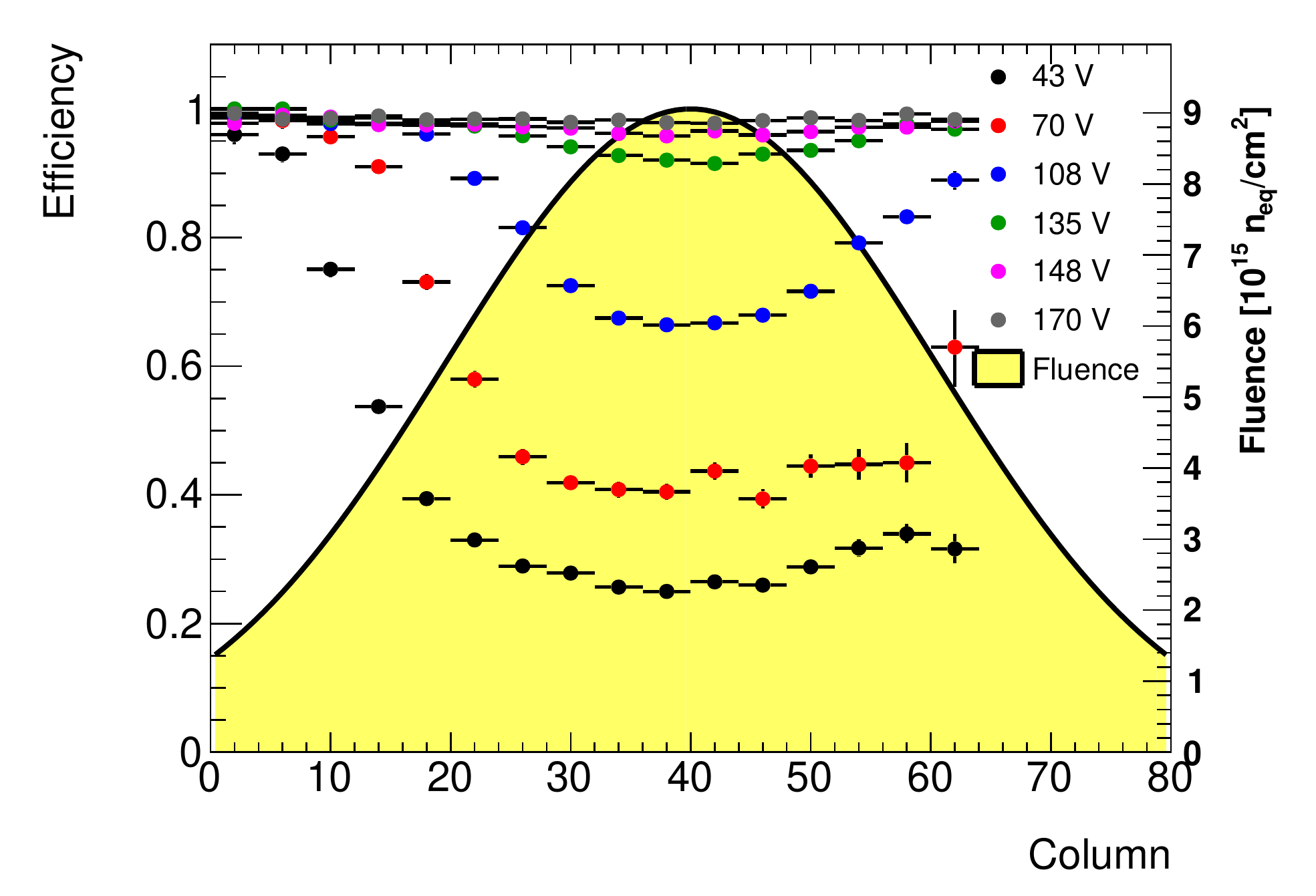}
	 \includegraphics[width=7.4cm]{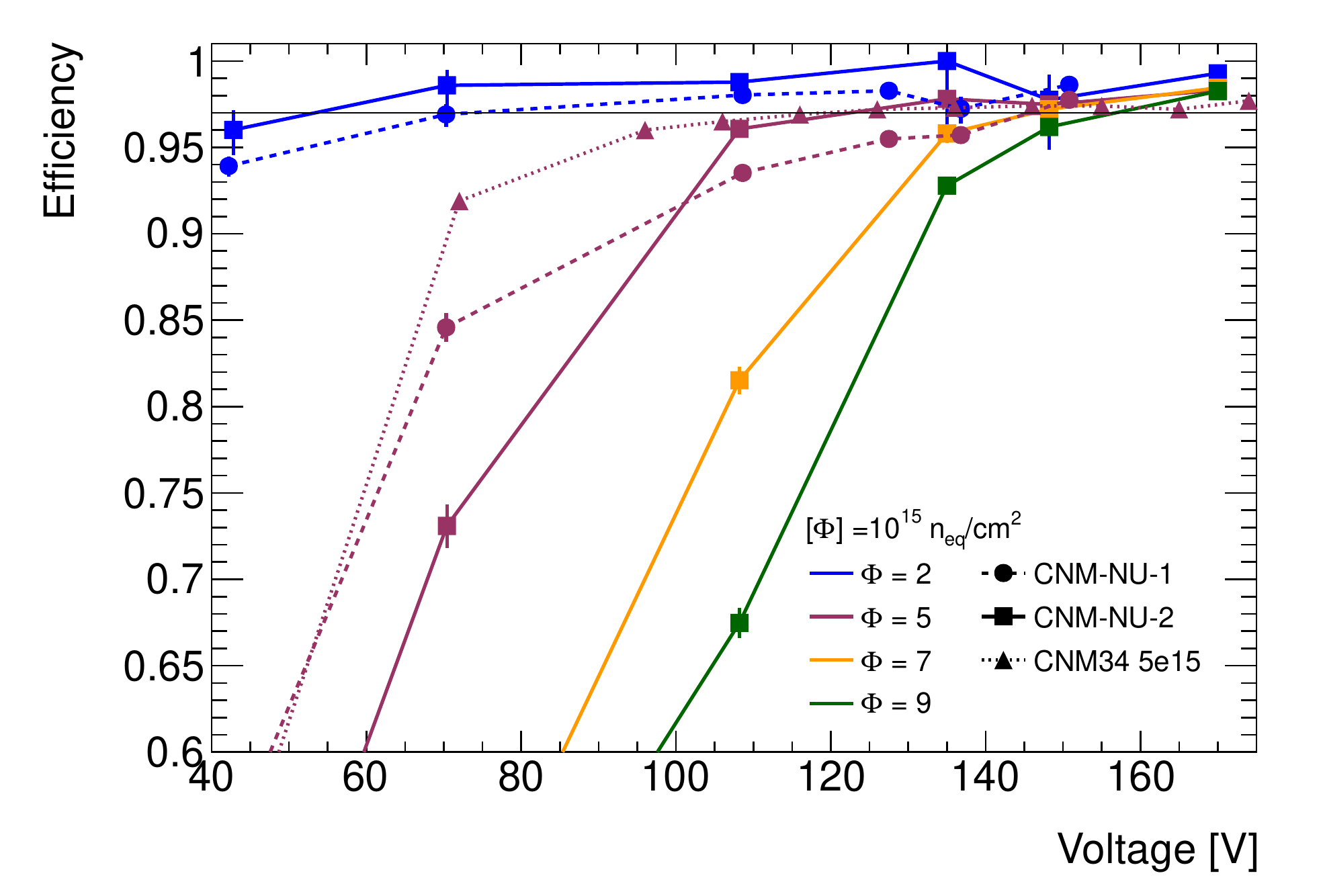}
	\includegraphics[width=7cm]{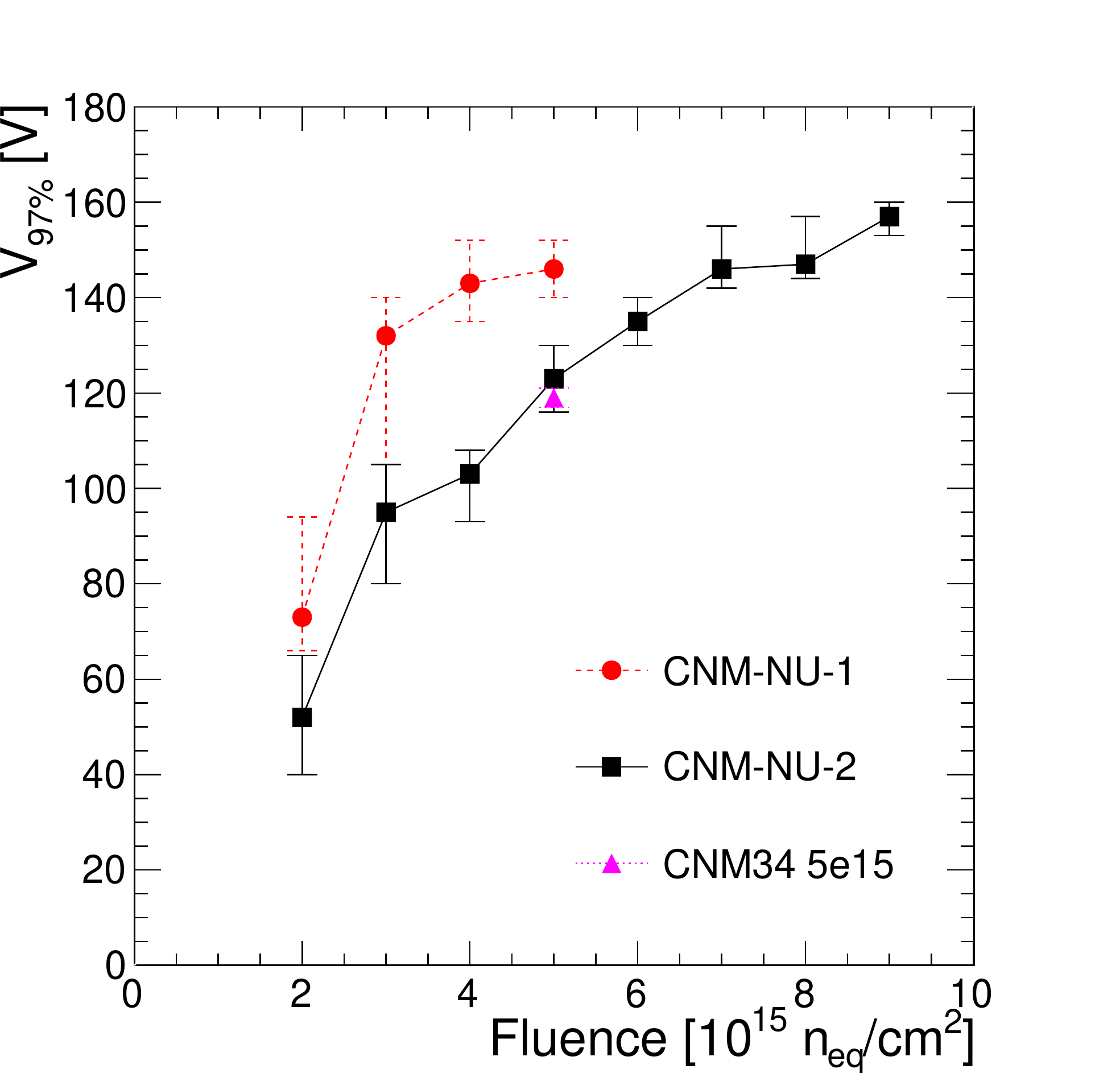}
	\caption{Top left: Hit efficiency as a function of pixel column number (250\,$\mu$m width) for different voltages for CNM-NU-2. Overlayed is the corresponding fluence distribution. Top right: Hit efficiency vs. voltage for different fluences for all devices. The 97\% benchmark efficiency is marked. Bottom: Voltage needed to reach 97\% efficiency as a function of fluence. The uncertainties are statistical only.}
	\label{fig:IBLgen}
\end{figure}

Two FE-I4 3D pixel detectors were irradiated at the CERN-PS IRRAD facility with 24\,GeV protons with a Gaussian beam profile of 12$\times$12\,mm$^2$ FWHM in December 2014. Hence, the samples show a non-uniform fluence over the detector area, ranging from about 15\% of the maximum fluence at the device edge to 100\% in the centre of the device (see figure~\ref{fig:IBLgen}, top left). Thus, it was possible to study a broad range of fluences on one single pixel detector. The fluence was normalised by gamma spectroscopy of a 5x5\,mm$^{2}$ aluminium piece in the centre of the pixel detector, which gave 5.2 and 8.4$\times10^{15}\,n_{eq}$/cm$^2$ for the devices CNM-NU-1 (5860-12-07) and CNM-NU-2 (6181-07-02), respectively. The corresponding fluences in the peak of the Gaussian beam profile are 5.6 and 9.1$\times10^{15}\,n_{eq}$/cm$^2$, respectively. The devices were annealed for one week at room temperature.
%(table~\ref{} shows the relation between the peak, mean and measured 5x5\,mm$^{2}$ fluence). 

These devices were measured in beam tests at the CERN SPS H6 beam line with 120\,GeV pions in July and September 2015 at 0$^{\circ}$ between the DUT sensor normal and the beam axis. They were tuned to a threshold of 1.5\,ke$^{-}$ and measured at a set temperature of -30$^{\circ}$C, corresponding to about -25$^{\circ}$C on sensor. A EUDET-type telescope made of six MIMOSA planes was used to provide reference tracks~\cite{bib:EUDET}. Figure~\ref{fig:IBLgen} (top left) shows the measured hit efficiency as a function of pixel column number (250\,$\mu$m width) averaged over a central band of 50 pixel rows for CNM-NU-2, measured at different voltages. Overlayed is the Gaussian fluence distribution. It can be seen that there is an efficiency minimum corresponding to the maximum fluence at the centre of the device, as expected. In fact, this minimum was in turn taken to fine-align the position of the fluence peak. From this plot, the voltage dependence of the efficiency was extracted for a number of different fluences (see figure~\ref{fig:IBLgen}, top right). The two devices agree reasonably well with each other, although CNM-NU-1 seems to have a general about 2\% lower plateau efficiency. At 5$\times10^{15}\,n_{eq}$/cm$^2$, they also agree with a device irradiated uniformly with 23\,MeV protons at Karlsruhe during the IBL qualification phase (CNM34)~\cite{bib:IBLprototypes}, which was remeasured here under the same conditions. This gives confidence that the method of the non-uniform irradiation and its calibration work well. Figure~\ref{fig:IBLgen} (bottom) shows the voltage $V_{97\%}$ at which the efficiency benchmark of 97\% is reached (linearly interpolated between measurement points), which is considered to be indicative for a suitable operational voltage. For the IBL target fluence of 5$\times10^{15}\,n_{eq}$/cm$^2$, $V_{97\%}$=120--145\,V is obtained, but also for the highest achieved fluence of 9$\times10^{15}\,n_{eq}$/cm$^2$, which is relevant for the inner pixel layers of the HL-LHC detectors, $V_{97\%}$ is found to be only moderately higher with about 157\,V. It should be noted that higher efficiencies and lower operation voltages are expected for the HL-LHC pixel detector due to a planned lower threshold of 1\,ke$^{-}$ and smaller electrode distances (see section~\ref{sec:HL-LHC}). Also tilting the sensor with respect to the beam is expected to further improve the efficiency~\cite{bib:IBLprototypes}.

%\begin{table}[htb]
			%\centering
			%\caption{Operational Voltage (where hit efficiency $>97$\%) for different fluences.}
			%\label{tab:Vop}
			%\begin{tabular}{|l|c|c|c|c|}
			%
%\hline												
	%&	\multicolumn{4}{|c|}{Fluence [$10^{15}\,n_{eq}$/cm$^2$]}	\\
%\cline{2-5}												
%Device	&	2	&	5	&	7	&	9			\\
%\hline												
%%6181-07-02	&	55\,V	&	123\,V	&	145\,V	&	157\,V	\\
%%CNM34	      &	--	  &	114\,V	&	--	    &	--	\\
%6181-07-02	&	55\,V	&	125\,V	&	145\,V	&	160\,V	\\
%CNM34	      &	--	  &	115\,V	&	--	    &	--	\\
%\hline												
%
%
%\end{tabular} 
	%\end{table}

%\subsection{Uniformly neutron-irradiated FE-I3 3D detectors}
%Whereas non-uniform irradiation gives the opportunity to study the efficiency for a large range of fluences, it is not ideal to study the leakage current and power dissipation since at such high fluences no current plateau is observed and hence no linear scaling of the current with fluence is reached for a fixed voltage. 
% They have a larger pixel size of 50x400\,$\mu$m$^{2}$, but due to three 3D electrodes per pixel, $L_{el}$=71\,$\mu$m is similar to the FE-I4. 

In addition, FE-I3 devices were irradiated with neutrons at JSI Ljbuljana to fluences of 0.5, 1.0, 1.5 and 2.0$\times10^{16}\,n_{eq}$/cm$^2$. The leakage current was measured after one week of annealing at room temperature in a climate chamber at -25$^{\circ}$C, the planned operation temperature for the ATLAS HL-LHC pixel detector. At the operation voltages obtained from the FE-I4 beam-test measurements above (conservatively taken or extrapolated as 145\,V and 180\,V for 0.5 and 1.0$\times10^{16}\,n_{eq}$/cm$^2$, respectively), the corresponding power dissipation was determined as 7 and 15\,mW/cm$^2$ for 0.5 and 1.0$\times10^{16}\,n_{eq}$/cm$^2$, respectively. The 1.0$\times10^{16}\,n_{eq}$/cm$^2$ device could be also measured in direct contact with a cold chuck, providing improved heat dissipation, which reduced the power dissipation to 12\,mW/cm$^2$. These values are considerably lower than for planar devices~\cite{bib:planarPower}. More details are presented in reference~\cite{bib:Iworid2016}.

%To obtain uniform irradiation for such studies and to minimise the damage of the readout chip by ionising dose during irradiation, reactor neutrons were chosen. 
%3D pixel detectors were irradiated at the TRIGA reactor at JSI Ljubljana with neutrons up to fluences of 0.5, 1.0, 1.5 and 2.0$\times10^{16}\,n_{eq}$/cm$^2$. However, tantalum in the FE-I4 chip leads to a large activation during irradiation, requiring long cool-down times, so that the FE-I3 chip was chosen instead. Figure~\ref{} shows the leakage current measured at -25$^{\circ}$C after one week of annealing at room temperature. The pixel detectors were typically measured assembled on a PCB in the climate chamber. The devices at 0.5 and 1.0$\times10^{16}\,n_{eq}$/cm$^2$ were in addition also measured detached from the PCB on a cold chuck for a better thermal contact with the cooling medium, which is more realistic in view of applications in the HL-LHC experiments. As can be seen, this reduces the current by FIXME. It can be seen that the current is higher for larger fluences, as expected, but no plateau is reached. For two devices irradiated to the same fluence, similar curves are obtained. The leakage current at the operational voltage as found from the measurements above for 0.5 and 1.0$\times10^{16}\,n_{eq}$/cm$^2$ (conservatively taken as 140 and 180\,V here) is even slightly below the model value when using the current damage parameter $\alpha=4.5\times10^{-17}$\,A/cm (and then scaled down to -25$^{\circ}$C). 

\section{Development of dedicated 3D pixel detectors for the HL-LHC}
\label{sec:HL-LHC}

Although already the results of the existing IBL/AFP generation of 3D sensors in terms of radiation hardness are very promising, a new generation of 3D pixel detectors needs to be developed for the HL-LHC with smaller pixel sizes to cope with the increased occupancy. A new, radiation-hard front-end chip with a threshold of 1\,ke$^-$ is being developed by the CERN RD53 collaboration, which can be matched to 50$\times$50 or 25$\times$100\,$\mu$m$^{2}$ sensor pixel size. With these reduced cell sizes, already the minimal configuration of 3D columns (1E, \ie one junction column per pixel) has a largely reduced inter-electrode distance compared to the IBL/AFP generation, \eg only 35\,$\mu$m for 50$\times$50\,$\mu$m$^{2}$ 1E (see figure~\ref{fig:3Dgeometry}). Thus, even less trapping and lower operational voltages are expected.

\begin{figure}[hbtp]
	\centering
	\includegraphics[width=15cm]{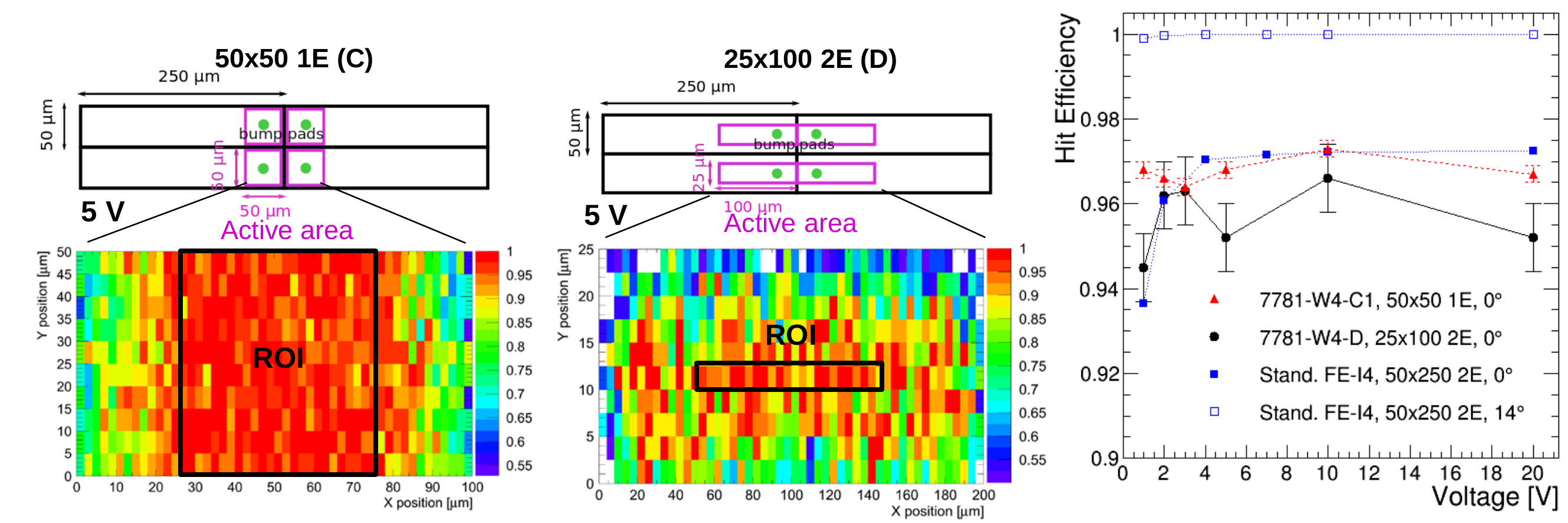}
	\caption{Top: sketch of 50$\times$50\,$\mu$m$^{2}$ (left) and 25$\times$100\,$\mu$m$^{2}$ (centre) sensor pixels matched to the larger FE-I4 chip pixels. Bottom: the corresponding hit-efficiency maps for 2 horizontally neighbouring sensor pixels at 5\,V. Right: average hit efficiencies in ROI vs. voltage for 50$\times$50 and 25$\times$100\,$\mu$m$^{2}$ pixel geometries at 0$^{\circ}$, compared to a standard FE-I4 at 0 and 14$^{\circ}$.}
	\label{fig:smallPitch}
\end{figure}

At CNM, a new 3D sensor run (7781) with small 3D cell sizes was carried out as CERN RD50 project and finished in December 2015. Since the RD53 readout chip is still under development, the small 3D sensor pixels are matched to the existing FE-I4 readout chip (see figure~\ref{fig:smallPitch} top left and centre sketches): each 50$\times$250\,$\mu$m$^{2}$ FE-I4 chip pixel cell contains five 50$\times$50\,$\mu$m$^{2}$ 1E or 25$\times$100\,$\mu$m$^{2}$ 2E sensor pixels, so that only 20\% of the sensor pixels can be connected to a front-end channel and be read-out. The remaining 80\% insensitive sensor pixels are shorted to ground to be at the same potential as the ones being read-out. 
%In addition, also a version with five 25$\times$100\,$\mu$m$^{2}$ 1E 3D unit cells shorted together to one large 25$\times$500\,$\mu$m$^{2}$ 5E pixel. 
Also strips and pad diodes of various geometries are included in the run. Five wafers finished the run and underwent electro-plate copper or electro-less gold Under-Bump Metalisation (UBM) at CNM. Several wafers broke after production. Eight devices were bump-bonded to the FE-I4 readout chip, assembled on readout boards and characterised at IFAE. Although some of them showed areas of disconnected bumps due to a poor UBM, all were suitable for testing. The breakdown voltage was typically between 15 and 40\,V, the devices could be tuned, and charge collection studies with a $^{90}$Sr source were successfully performed. More details of the run and initial laboratory characterisations are described in reference~\cite{bib:Iworid2016}. The yield and the breakdown voltages of this first small-pixel prototype run are not ideal, but in the meanwhile large process improvements have been achieved at CNM to solve this, as already demonstrated by a new AFP run~\cite{bib:AFPproduction}.

In May and June 2016, beam tests were performed on unirradiated small-pixel devices at the CERN SPS H6 beam line with 120\,GeV pions. Due to initial technical problems with EUDET-type telescopes, a custom-made 3D FE-I4 telescope was used. It consisted of two equal arms up- and down-stream of the devices under test (DUTs). Each arm was made of two modules with the short 50\,$\mu$m pixel direction in $y$ and one module rotated with the short pixel direction in $x$. Each module was tilted by 14--15$^\circ$ with respect to the short pixel direction to increase the position resolution to about 6\,$\mu$m per plane in that direction~\cite{bib:AFPbeamTests}. Hence, the track resolution at the DUT position is expected to be about 3\,$\mu$m in $y$ and 4\,$\mu$m in $x$ (neglecting the long pixel direction). The DUT sensor normal was placed at 0$^{\circ}$ with respect to the beam axis.

Figure~\ref{fig:smallPitch} (bottom) shows the in-pixel hit-efficiency map for reconstructed tracks restricted to the active small-pixel sensor area over two horizontally neighbouring sensor pixels. For the determination of an average efficiency, due to the telescope resolution smearing, the region of interest (ROI) was even further reduced to the central 50$\times$50\,$\mu$m$^{2}$ in the two-neighbouring-sensor-pixel area for the 50$\times$50\,$\mu$m$^{2}$ geometry and a central 2.5$\times$100\,$\mu$m$^{2}$ area for the 25$\times$100\,$\mu$m$^{2}$ device. Figure~\ref{fig:smallPitch} (right) shows the hit efficiency in this ROI as a function of bias voltage. Already at 1--2\,V the small-pixel devices reach their plateau efficiencies of 96--97\%, similar to a medium-quality standard reference FE-I4 device, which however needs 4\,V for full efficiency due to larger $L_{el}$ and hence later full depletion. High-quality class standard FE-I4 devices can even reach up to 99\% at 0$^{\circ}$~\cite{bib:IBLprototypes}, which might be achievable as well for small-pixel devices with improved processing. However, at the example of the medium-quality standard FE-I4 detector, it is shown that tilting by 14$^{\circ}$ greatly improves the efficiency to 99.9\% due to minimising the influence of low-efficiency regions from the 3D columns or low-field areas. Such behaviour is also expected from the small-pixel devices, which will be verified in further beam tests.

First measurements have been performed on irradiated 50$\times$50 and 25$\times$100\,$\mu$m$^{2}$ strip detectors of this run~\cite{bib:Iworid2016}. Initial results hint at increased leakage currents for smaller electrode distances, but for all devices the power dissipation is below 25\,mW/cm$^2$ up to 150\,V at $10^{16}\,n_{eq}$/cm$^2$. It is still under investigation whether this is an artifact of this first run with non-ideal leakage currents already before irradiation or a real effect of enhanced charge multiplication in higher fields at smaller electrode distances. In any case, the operation voltage is expected to decrease, which has a compensating effect. Further irradiations and beam tests are carried out to clarify this.

\section{Conclusions and outlook}
\label{sec:conclusions}

3D silicon pixel detectors have been investigated as candidates for the innermost pixel layers for the HL-LHC upgrade of the ATLAS detector, and have demonstrated a good performance. 

Firstly, the existing IBL/AFP 3D generation was studied at HL-LHC fluences of about $10^{16}\,n_{eq}$/cm$^2$. Non-uniformly proton-irradiated standard FE-I4 detectors have demonstrated an efficiency of 97\% at only 157\,V for a fluence of $9\times10^{15}\,n_{eq}$/cm$^2$ and 1.5\,ke$^-$ threshold. The power dissipation of neutron-irradiated FE-I3 detectors was found to be 12--15\,mW/cm$^2$ at -25$^{\circ}$C for a fluence of $10^{16}\,n_{eq}$/cm$^2$, which is significantly smaller than for planar pixel detectors. Hence, already the IBL/AFP 3D generation seems radiation-hard enough for the use in the HL-LHC. In the meanwhile, IBL-type devices have been irradiated to even higher fluences and measured in beam tests. The analysis is on-going.

%However, due to higher occupancies, smaller pixel sizes are mandatory for HL-LHC pixel detectors, which is expected to even further improve radiation hardness. 
A first 3D sensor production was carried out at CNM with sensor pixel sizes of 50$\times$50 and 25$\times$100\,$\mu$m$^{2}$ as planned for the HL-LHC. Although the yield and breakdown voltages are not ideal, the non-irradiated devices have demonstrated a good performance in laboratory tests and reached about 97\% efficiency in beam tests already at 1--2\,V, similar to medium-quality class standard FE-I4s. The efficiency is expected to improve by tilting, which is still being analysed. Furthermore, strip, pad and pixel detectors of this production have been already irradiated at different facilities to HL-LHC fluences and measured in beam tests. The analysis is on-going.

Further 3D productions are on-going at CNM: one similar to the small-pixel run presented, but with improved processes in order to increase the yield and breakdown voltages, as has been already demonstrated by a new AFP 3D production~\cite{bib:AFPproduction}; another run with the same mask, but as a single sided 3D process allowing the production of 70--150\,$\mu$m thin wafers; and lastly a run with small pixel structures compatible with the RD53 readout chip. Also other vendors like FBK, Italy, and SINTEF, Norway, have productions on-going for HL-LHC 3D pixel detectors.

\acknowledgments %JINST
%\section*{Acknowledgments} %ATLASnote
The authors wish to thank A.\,Rummler, M.\,Bomben and the other ATLAS ITk beam test participants for great support and discussions at the beam tests; also to F.\,Ravotti and G.\,Pezzullo (CERN IRRAD), as well as V.\,Cindro and I.\,Mandic (JSI Ljubljana) for excellent support for the irradiations. This work was partly performed in the framework of the CERN RD50 collaboration.
This work was partially funded by: the MINECO, Spanish Government, under grants FPA2013-48308-C2-1-P, FPA2015-69260-C3-2-R, FPA2015-69260-C3-3-R and SEV-2012-0234 (Severo Ochoa excellence programme) and under the Juan de la Cierva programme; the Catalan Government (AGAUR): Grups de Recerca Consolidats (SGR 2014 1177); and the European Union's Horizon 2020 Research and Innovation programme under Grant Agreement no. 654168.

%%%%%%%%%%%%%%%%%%%%%%%%%%%%%%%%%%%%%%%%%%%%%%%%%%%%%%%%%%%%%%%%%%%%%%%%%%%%%%%
% Bibliography
%%%%%%%%%%%%%%%%%%%%%%%%%%%%%%%%%%%%%%%%%%%%%%%%%%%%%%%%

%\bibliographystyle{atlasBibStyleWithTitle}
%\bibliography{Lange_bibliography}

%JINST bibliography:
\providecommand{\href}[2]{#2}\begingroup\raggedright\endgroup

\end{document}